\documentclass[a4paper,11pt]{article}
\usepackage{graphicx} 
\usepackage{amsmath,amssymb,slashed}
\usepackage[dvipsnames]{xcolor}
\usepackage{xspace}
\usepackage{subcaption}
\usepackage{siunitx, booktabs,array}

\usepackage{jcappub} 
\newcommand{\Neff}{\ensuremath{N_{\mathrm{eff}}}\xspace}
\newcommand{\DNeff}{\ensuremath{\Delta N_\mathrm{eff}}\xspace}
\newcommand{\lcdm}{\ensuremath{\Lambda{\mathrm{CDM}}}\xspace}

\title{InterACTing dark radiation models after ACT}

\date{\today}
\author[]{William Cvetko,}
\author[]{Melissa Joseph,}
\author[]{Gustavo Marques-Tavares}
\affiliation[]{Department of Physics and Astronomy, University of Utah, Salt Lake City, UT 84112, U.S.A.}

\abstract{
In this work we assess the implications of the Atacama Cosmology Telescope DR6 measurements for two interacting dark radiation scenarios previously shown to mitigate the Hubble tension. The first model, Wess--Zumino dark radiation (WZDR), features a mass threshold in the dark sector that induces a step-like reduction in the dark radiation abundance as the dark temperature evolves. The second model, new atomic dark matter (nuADaM), introduces dark radiation that remains coupled to a subcomponent of dark matter until shortly before matter--radiation equality. Earlier analyses using \emph{Planck} data demonstrated that these interactions significantly relax constraints on the dark radiation density and allow values of $H_{0}$ consistent with local distance-ladder determinations. Incorporating ACT DR6, which extends CMB measurements deep into the high-$\ell$ damping tail, we find that constraints on the additional radiation component tighten substantially in both scenarios, closing most of the parameter space that previously enabled higher values of $H_{0}$. We further analyze a generalized framework including both free-streaming and self-interacting dark radiation, and show that the resulting constraints are consistent with ACT’s findings for the limiting cases of purely free-streaming or purely self-interacting radiation. Overall, ACT DR6 significantly restricts interacting dark radiation as a solution to the Hubble tension.
}

\begin{document}

\maketitle

\section{Introduction}
The Hubble tension stands as one of the most significant challenges in cosmology: local distance ladder measurements yield $H_0 = 73.04 \pm 1.04$ km/s/Mpc \cite{Riess:2021jrx} while the CMB-inferred Planck value gives $H_0 = 67.4 \pm 0.5$ km/s/Mpc \cite{Planck:2018vyg}, a $\gtrsim 5\sigma$ discrepancy.\footnote{%
It is worth noting that the Chicago-Carnegie Hubble program collaboration finds a smaller tension~\cite{Freedman:2021ahq,Freedman:2024eph}.} 
This Hubble tension has motivated wide-ranging investigations into extensions of the standard cosmological model~(see Refs.~\cite{Buen-Abad:2015ova,Lesgourgues:2015wza,Buen-Abad:2017gxg,Zhao:2017cud,DiValentino:2017gzb,Poulin:2018cxd,Smith:2019ihp,Lin:2019qug,Alexander:2019rsc,Agrawal:2019lmo,Escudero:2019gvw,Berghaus:2019cls,Vagnozzi:2019ezj,Ye:2020btb,Das:2020xke,RoyChoudhury:2020dmd,Brinckmann:2020bcn,Krishnan:2020vaf,Seto:2021xua,Ye:2021iwa,Niedermann:2021vgd,Aloni:2021eaq, Dainotti:2021pqg,Odintsov:2022eqm,Berghaus:2022cwf,Schoneberg:2022grr,Joseph:2022jsf,Buen-Abad:2022kgf,Rezazadeh:2022lsf,Brinckmann:2022ajr,Wang:2022bmk,Bansal:2022qbi,Buen-Abad:2023uva,Sandner:2023ptm,Niedermann:2023ssr,Hughes:2023tcn,Greene:2024qis,Allali:2024anb, Co:2024oek,Cho:2024lhp,Simon:2024jmu,DeSimone:2024lvy,Montani:2024ntj,Buen-Abad:2024tlb,Chang:2025uvx,Garny:2025kqj,GarciaEscudero:2025orc} for a partial list of proposals, in addition to the models reviewed in~\cite{DiValentino:2021izs,Schoneberg:2021qvd, Abdalla:2022yfr,Escudero:2022rbq,Poulin:2023lkg,Khalife:2023qbu}), many of which proposing new dynamics in the dark sector.

\begin{figure}[h]
\centering
\includegraphics[width=0.9\textwidth]{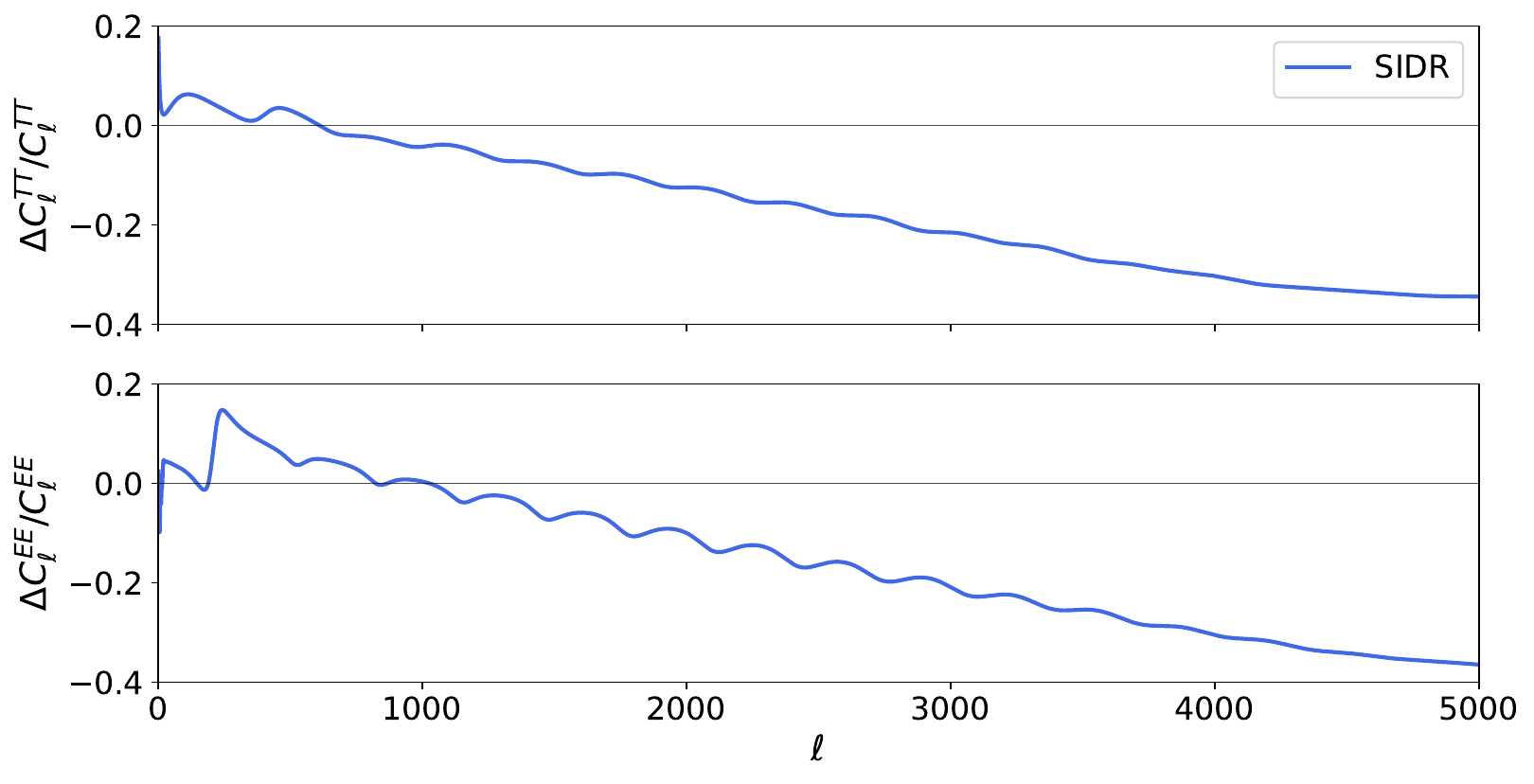}
\caption{CMB residuals of SIDR compared to \lcdm, using the best fit points from a fit to Planck+SH0ES from Ref~\cite{Aloni:2021eaq}.}
\label{fig:planck-cmb-residuals}
\end{figure}

Among the proposed solutions, a well motivated class of beyond the standard model scenarios are dark sectors with additional relativistic particles in the early universe (dark radiation). In these scenarios, the dark radiation increases the expansion rate prior to recombination, reducing the sound horizon and, therefore, raising the inferred $H_0$ from CMB data. The simplest extensions, $\Lambda$CDM+$\Delta N_{\rm eff}$ with additional free-streaming radiation or self-interacting radiation (SIDR) were previously found to accommodate $\Delta N_{\rm eff} \sim 0.2$ (free-streaming) and $N_{\rm idr} \sim 0.3$ (self-interacting) when fit to Planck~\cite{Schoneberg:2021qvd}. However, the recent ACT DR6 analysis of extended models~\cite{ACT:2025tim} found strong constraints on extra dark radiation from high-precision temperature and polarization data at $\ell > 1000$ combined with large scale data from Planck, with constraints $\Delta N_{\rm eff} < 0.14 $ (95\% CL, P-ACT) and a similar bound for SIDR $N_{\rm idr} < 0.114 $ (95\% CL, P-ACT). This result reflects the increased sensitivity of ACT DR6 at high $\ell$, where the presence of extra radiation leads to a suppression of power due to increased Silk damping~\cite{Smith:2025zsg}. In Fig.~\ref{fig:planck-cmb-residuals}, we show the residuals between the best fits for SIDR and \lcdm found in Ref.~\cite{Aloni:2021eaq}, which included the SH0ES likelihood, extended to $\ell \leq 5000$. One can clearly see the extra suppression at $\ell \gtrsim 1000$ from having $N_{\rm idr} = 0.47$, which was not strongly constrained by Planck but can be probed at high precision by ACT DR6 data.
 
Previous work had found that models with non-trivial dynamics for dark radiation allowed for even larger values of \DNeff, and thus of $H_0$~\cite{Aloni:2021eaq,Allali:2023zbi,Schoneberg:2023rnx,Buen-Abad:2024tlb}. This is due to a combination of different effects. The fluid-like behavior of SIDR leads to a slower decay of the gravitational potential perturbations which leads to slightly more power at small scales, in addition to shifting the large $\ell$ peaks in the opposite direction to free-streaming species. This is why \DNeff could be larger for SIDR models. Furthermore, models with time dependent dynamics, such as the Wess-Zumino Dark Radiation model~\cite{Aloni:2021eaq} and the new atomic dark matter model~\cite{Buen-Abad:2024tlb}, had interactions that modified the evolution of dark radiation perturbations at a transition time close to equality. This leads to $\ell$ dependent effects, which combined with changes to other parameters allows for a larger density in radiation without significant suppression of power on the scales measured by Planck.
This motivates our central question: do these theoretically motivated models of interacting dark radiation still allow for significant energy density in dark radiation when ACT DR6 data are included, or are they also severely constrained? 

In this work, we present an updated analysis of WZDR and nuADaM models using ACT DR6 data combined with complementary cosmological probes including Planck, CMB lensing, DESI DR2 BAO measurements, and Type Ia supernovae. We find that despite analysis with Planck data allowing much larger values for \DNeff compared to simple models, both WZDR and nuADaM face constraints from ACT DR6 comparable to those on simpler dark radiation models. The scale-dependent signatures prove insufficient to evade the fundamental tension between extra radiation density and the precisely measured high-$\ell$ damping tail, significantly impacting their potential as a solution to the Hubble tension.

\section{Models} \label{sec:models}

The most well-known extension, $\Lambda$CDM+$\Delta N_{\rm eff}$, adds free-streaming radiation that behaves identically to Standard Model neutrinos in its perturbative evolution. An alternative scenario is self-interacting dark radiation (SIDR), where the additional radiation forms a perfect fluid with $w = c_s^2 = 1/3$. The fluid-like behavior of SIDR leads to different effects on the CMB, at the perturbation level, compared to free-streaming radiation~\cite{Chacko:2003dt,Baumann:2015rya,Brust:2017nmv}. The combined model analyzed in Section \ref{sec:results} alters the radiation density in both free-streaming and self-interacting radiation, possibly corresponding to a model in which energy is injected into the SM photon bath at early times along with a interacting dark radiation component. In addition to the combined model, we consider two interacting dark radiation models that undergo a transition at a redshift close to matter-radiation-equality, which leads to $\ell$-dependent effects in the CMB power spectra.


\subsection{WZDR Model}
The WZDR model consists of an interacting dark sector composed of two particles: a massive complex scalar ($\phi$) and a massless Weyl fermion ($\psi$) interacting through a Yukawa coupling. 
\begin{equation}
{\cal L}_\text{WZ} = \frac{1}{2} m_\phi^2 \phi^2+  \lambda \, \phi \, \psi^2 + \lambda^2 \, (\phi^* \phi)^2
~.
\end{equation}
The cosmological evolution of the dark radiation follows from the specified interactions. At early times, when the dark sector temperature $T_d \gg m_\phi$, both the fermion and the scalar are relativistic, contributing an effective number of neutrino species $N_{\rm UV}$ to the radiation budget. The dark sector is assumed to be populated after Big Bang Nucleosynthesis (BBN) through some unspecified mechanism (such as freeze-in from Standard Model neutrinos), so BBN predictions remain unaffected. As the universe expands and cools, the temperature eventually drops below the scalar mass at redshift $1 + z_t \approx m_\phi / T^0_{d}$, where $T^0_{d}$ is the dark sector temperature today. At this transition epoch, the scalars become non-relativistic and begin to decay. We consider couplings such that decay occurs in equilibrium and the effective four-fermion interaction $\lambda^2 \psi^4 / m_\phi^2$ is large enough for the fermions to remain in thermal equilibrium until recombination. Because dark radiation maintains chemical and kinetic equilibrium through self-interactions, the entropy from the scalars is efficiently deposited into the massless fermion component. This entropy transfer occurs over approximately one decade in redshift and causes the effective number of relativistic degrees of freedom to increase from $N_{\rm UV}$ to $N_{\rm IR} = (15/7)^{1/3} N_{\rm UV}$, where the factor $(15/7)^{1/3}$ follows from counting the bosonic and fermionic degrees of freedom before and after the transition.

The model has two free parameters: $N_{\rm IR}$ quantifying the late-time radiation density and $\log_{10}(z_t)$ specifying the transition redshift. We fix the ratio $N_{\rm IR}/N_{\rm UV} = (15/7)^{1/3}$ as predicted by the WZDR model consisting of a complex scalar and a Weyl fermion. 
An extension of this model was proposed in Ref.~\cite{Joseph:2022jsf}, which included feeble interactions between the WZDR radiation and dark matter to alleviate the $S_8$ tension. Given that the tension has become negligible in more recent analysis~\cite{Sailer:2024jrx,Qu:2024sfu,Wright:2025xka}, we focus our analysis on the original model which contains fewer parameters.

\subsection{nuADaM Model}

The new atomic dark matter model (nuADaM)~\cite{Buen-Abad:2024tlb}, extends the atomic dark matter model by considering a richer dark radiation component, which remains interacting at all relevant cosmological epochs. In this model a sub-component of dark matter is made of two particle species, one heavier (the dark proton $p'$) and one lighter (the dark electron, $e'$). These particles are charged under a dark $U(1)$ gauge interaction mediated by the dark photon, with dark fine structure constant given by $\alpha_d$. The relic abundance of these particles is set by an asymmetry, similar to the baryon asymmetry, such that the dark proton has positive charge and dark electron has negative charge. At high temperatures $T \gg \alpha^2_d m_{e'}$, the dark charged particles are ionized and scatter frequently with the dark gauge boson, while at low temperatures $T \ll \alpha_d^2 m_{e'}$, they combine into dark atoms and decouple from the dark radiation bath. The dark radiation consists of the aforementioned dark photon, another massless dark vector $X$ which mixes with the dark photon, plus massless fermions $\nu'_i$ charged under this second vector $X$. Through kinetic mixing with X, the dark photon can interact with both $\nu'$ and $X$ at all relevant times, ensuring that dark radiation remains a self-interacting fluid during all relevant times. The lagrangian describing this dark sector is given by:
\begin{equation}
\begin{aligned}
     \mathcal{L}_{\text{nuadam}}  = & -\frac{1}{4}F'^{\mu \nu}F'_{\mu \nu} + \bar p' ( i \slash  \!\!\!\! D - m_{p'} ) p' +  \bar e' ( i \slash  \!\!\!\! D - m_{e'} ) e'   -\frac{1}{4} X^{\mu \nu} X_{\mu \nu}  \\
     & - \frac{\epsilon}{4} F'^{\mu \nu} X_{\mu \nu} + \sum\limits_{j=1}^{N_f} \bar \nu'_{j} \, i \, \slash \!\!\!\! D \nu'_{j} \,
\end{aligned}
\end{equation}
where $F^{\mu \nu}$ is the field strength of the dark photon.

While there are three interactions associated with the model and two mass scales, in the regime of interest there are only three parameters that are relevant for the cosmological signatures of this scenario: the energy density in dark radiation $N_{\rm idr}$, the fraction of the dark matter energy density in atomic dark matter ($f_\text{idm}$) and redshift when dark recombination begins, $a_t$. Instead of working directly with $a_t$, which depends on the dark sector temperature (and thus on $N_{\rm idr}$), we fix $\alpha_d = 0.01$ and $m_{p'} = 1 \, \text{GeV}$ and use the ratio $m_{e'}/m_{p'}$ as the parameter controling when the transition occurs.

\section{Data and Methodology}\label{sec:data&m}

Our analysis utilizes the Atacama Cosmology Telescope Data Release 6 (ACT DR6) in combination with complementary cosmological probes. The release provides temperature and polarization power spectra (TT, TE, EE) over the multipole range $600 < \ell < 6500$, with white noise levels improved by factors of two (temperature) and three (polarization) relative to Planck in most of the range of overlapping $\ell$. 

Our analysis employs the same baseline datasets as the ACT DR6 cosmology analysis~\cite{ACT:2025fju}, combined with additional low-redshift probes to constrain late-time expansion history. We define dataset combinations using nomenclature consistent with the ACT DR6 baseline paper:
\textbf{p-ACT}:
\begin{itemize}
    \item \textbf{ACT DR6}: CMB power spectra (TT, TE, EE) from $600 < \ell < 6500$ using the \texttt{ACT-lite} likelihood~\cite{ACT:2025fju}. 
    \item \textbf{Planck 2018}: We use a \texttt{Planckcut} likelihood incorporating TT power spectra at $\ell < 1000$ and TE/EE at $\ell < 600$ from the PR3 release~\cite{Planck:2018vyg}, minimizing overlap with ACT while preserving Planck's cosmic-variance-limited constraints at large scales. We include the \texttt{Sroll2} low-$\ell$ EE likelihood ($\ell < 30$) for the reionization optical depth.
\end{itemize}
\textbf{P-ACT-L} (+ CMB Lensing)
\begin{itemize}
    \item \textbf{CMB Lensing}: We combine ACT DR6 lensing~\cite{ACT:2023kun,ACT:2023dou} with Planck PR4 NPIPE lensing~\cite{Carron:2022eyg}.
\end{itemize}
\textbf{P-ACT-LB} (+ BAO):
\begin{itemize}
    \item \textbf{BAO}: We include DESI Data Release 2 (DR2) measurements~\cite{DESI:2025zgx} spanning three years of observations across five tracer populations: Bright Galaxy Survey ($z < 0.4$), Luminous Red Galaxies ($0.4 < z < 1.1$), Emission Line Galaxies ($1.1 < z < 1.6$), Quasars ($0.8 < z < 2.1$), and Lyman-$\alpha$ forest ($2.0 < z < 4.2$).\footnote{Note that, while most of the data used in our analysis is the same as used in Ref.~\cite{ACT:2025tim}, we are using the updated DR2 results from DESI instead of DR1.}
\end{itemize}
\textbf{P-ACT-LBS} (+ Supernovae):
\begin{itemize}
    \item \textbf{Type Ia Supernovae}: Pantheon+ compilation~\cite{Scolnic:2021amr} includes 1701 light curves from 1550 SNe Ia spanning $0.001 < z < 2.26$.
\end{itemize}
We additionally include the following datasets labeled \textbf{SH0ES} and \textbf{$S_8$}
\begin{itemize}
    \item \textbf{$H_0$ Measurement}: SH0ES local distance ladder measurement of $H_0 = 73.04 \pm 1.04$ km/s/Mpc~\cite{Riess:2021jrx}, implemented as a Gaussian likelihood on the intrinsic brightness of the supernovae, $M_b = -19.253 \pm 0.027$. 
    \item \textbf{$S_8$ Measurement}: cosmic shear constraints from the completed Kilo-Degree Survey (KiDS)~\cite{Wright:2025xka}, implemented as a Gaussian likelihood - $S_8 = 0.815^{+0.016}_{-0.021}$.
\end{itemize}
We perform parameter estimation using modified versions of the CLASS Boltzmann code~\cite{Blas:2011rf} containing implementations of the WZDR and nuADaM models, interfaced with the Cobaya framework~\cite{Torrado:2020dgo} for Markov Chain Monte Carlo sampling. Minimization was done using Procoli algorithm~\cite{Karwal:2024qpt} in COCOA\footnote{\url{https://github.com/CosmoLike/cocoa}}. The WZDR implementation follows the approach of Ref.~\cite{Aloni:2021eaq}, while the nuADaM implementation\footnote{%
\url{https://github.com/ManuelBuenAbad/class_nuADaM}} is based on Ref.~\cite{Buen-Abad:2024tlb}. Model-specific parameters are varied with flat priors: for WZDR, $N_{\rm IR} \in [0, 5]$ and $\log_{10}(a_t) \in [-5.0, -4.0]$; for nuADaM, $\Delta N_{\rm idr} \in [0, 5]$, $\log_{10}(m_{e'}/m_{p'}) \in [-4.5, -3.5]$, and $r_{\rm twin} \in [0, 0.3]$. Convergence of the Markov chains is assessed using the Gelman-Rubin criterion, requiring $R-1 < 0.05$ for all parameters (for WZDR all the runs reached $R-1<0.01$).

\section{Results} \label{sec:results}

\begin{figure}[ht]
\centering
\includegraphics[width=0.45\textwidth]{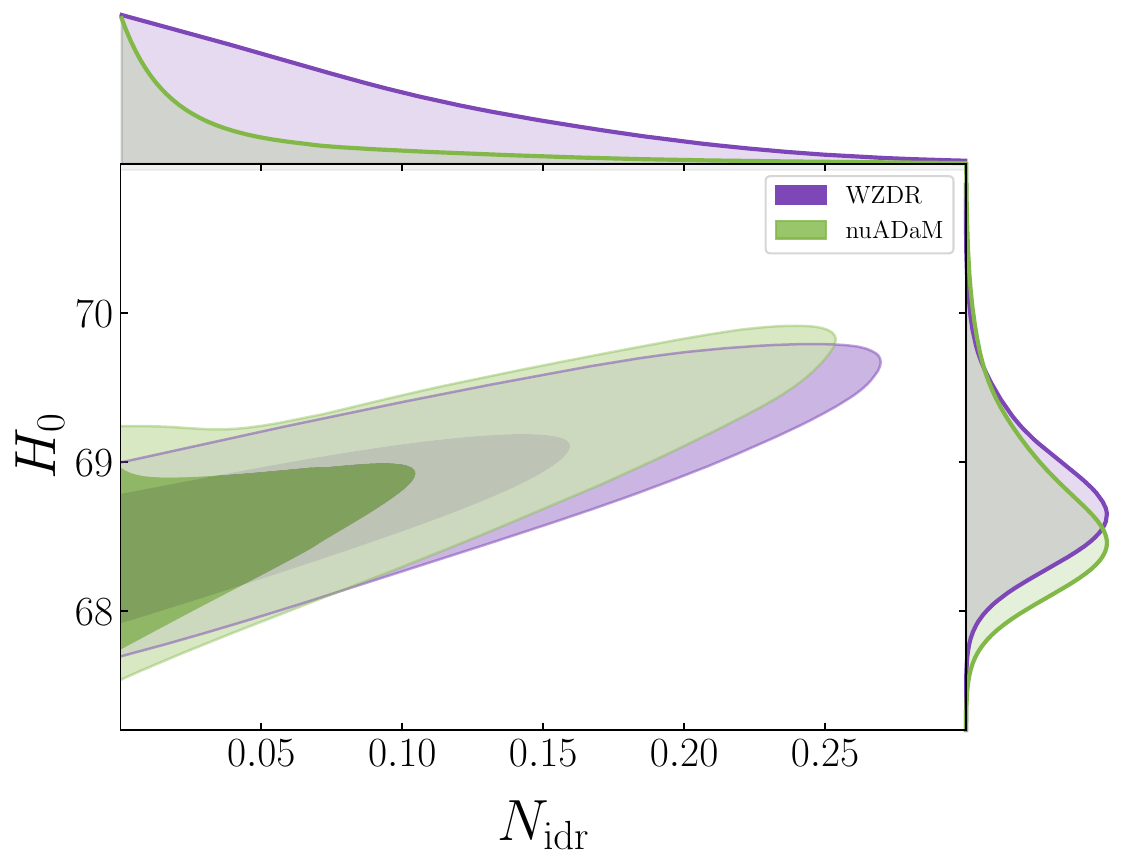}
\includegraphics[width=0.48\textwidth]{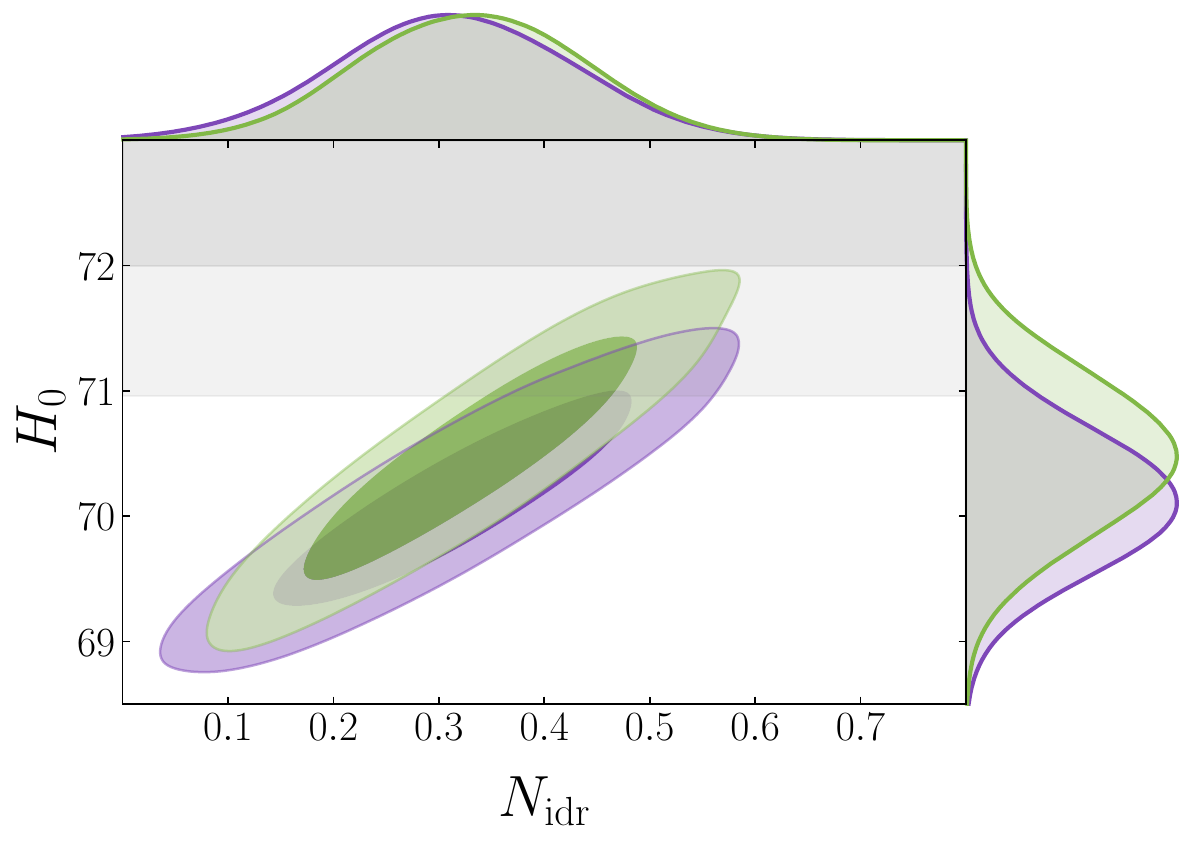}
\caption{{\bf Left figure:} $67\%$ and $95\%$ CL for $H_0$ and $N_\text{idr}$ for WZDR (purple) and nuADaM fit to {\bf P-ACT-LB} + $S_8$. {\bf Right figure:} $67\%$ and $95\%$ CL for $H_0$ and $N_\text{idr}$ for WZDR (purple) and nuADaM fit to {\bf P-ACT-LB} + $S_8$ + SH0ES. The gray band correspond to the $1\sigma$ and $2 \sigma$ regions for $H_0$ from the SH0ES collaboration~\cite{Riess:2021jrx}.}
\label{fig:Neff}
\end{figure}

First we present our results when fitting the models to the data described in the previous section, with and without the inclusion of the SH0ES likelihood. In Fig.~\ref{fig:Neff}, we see that the maximum values of $N_\text{idr}$ and $H_0$ at $95\%$ CL are significantly reduced from what previous studies found using Planck data~\cite{Aloni:2021eaq,Buen-Abad:2024tlb}, even when the data includes the SH0ES likelihood. The figure shows that for the fits including SH0ES, the tension reduces significantly, even if the $67\%$ CL only allows $H_0$ values that are much lower than what had been found in fits without ACT. The triangle plots with the posterior for all parameters are in Appendix~\ref{app:triangle}.

\begin{figure}[h]
\centering
\includegraphics[width=0.9\textwidth]{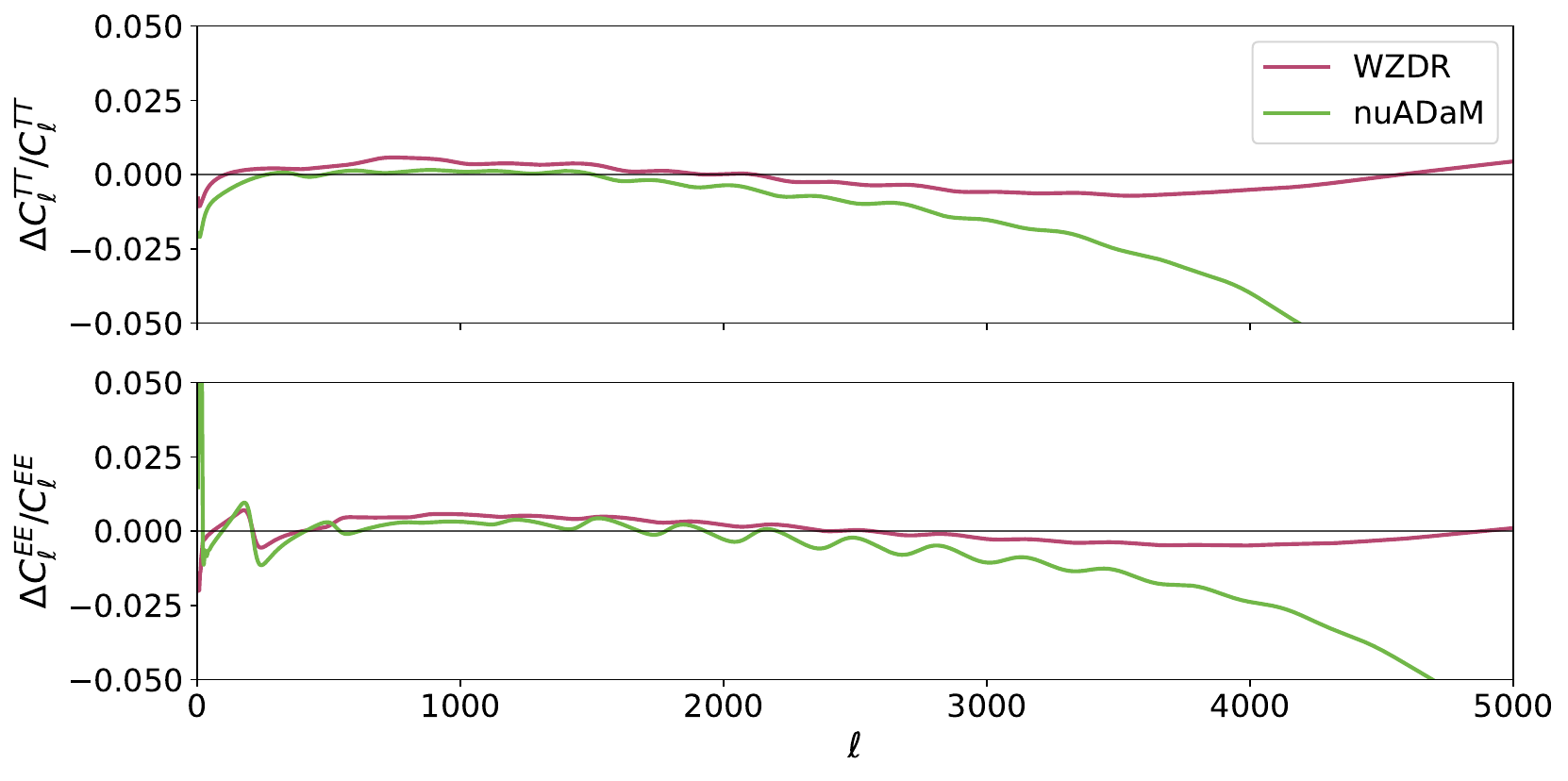}
\caption{CMB residuals of WZDR and NuADaM compared to \lcdm, using the best fit points in Table~\ref{tab:best-fit}.}
\label{fig:cmb-residuals}
\end{figure}

Table~\ref{tab:best-fit} shows the best fit of the models compared to \lcdm and their respective $\chi^2$, broken down by the contributions from each dataset. Note that the best fit $H_0$ for both WZDR and nuADaM are still $\gtrsim 2\sigma$ below the SH0ES value, whereas in previous analysis they were within $1\sigma$. Moreover, although both models exhibit an overall improvement in total $\chi^2$ relative to $\Lambda$CDM, this gain is driven exclusively by a reduced $H_0$ tension and is accompanied by a noticeable degradation in the CMB fit. These results indicate that the inclusion of ACT data substantially weakens the ability of these models to resolve the Hubble tension.\footnote{%
This is in line with the results of Ref.~\cite{Buen-Abad:2025bgd}, which analyzed a set of phenomenological dark matter-dark radiation models which has similar features to nuADaM and reached a similar conclusion about the inclusion of ACT data.} In order to understand what is the main driver in the constraints, we plot the TT and EE residuals between the best fit points from Table~\ref{tab:best-fit} in Fig.~\ref{fig:cmb-residuals}. We can see that both models predict a decrease in power at high $\ell$ due to the increased Silk damping compared to \lcdm. The results indicate that such a decrease is disfavored by ACT data, as was expected from the fact that the collaboration found that their best fit for variable \Neff found $N_\text{eff} = 2.86 \pm 0.13$~\cite{ACT:2025tim} and for \lcdm, $n_s = 0.9752 \pm 0.0030$~\cite{ACT:2025fju}, both of which correspond to more power at high $\ell$ than predicted from the best fits to Planck data~\cite{Planck:2018vyg}.

\begin{table}[ht]
\centering
\begin{tabular}{|l|c|c|c|}
\hline
Parameter & WZDR & nuADM & $\Lambda$CDM \\
\hline
$\Delta N_{\rm idr}$       & 0.3052   & 0.3620  & -- \\ 
$\log_{10} a_{\rm t}$      & -4.62   & --   & -- \\ 
$\log_{10}(m_{e'}/m_{p'})$
     & --  & -3.720  & -- \\ 
$r_{\rm twin}$             & --          & 0.0145  & -- \\ 
$100\,\theta_s$            & 1.041980    & 1.042412    & 1.041849 \\ 
$n_s$                      & 0.9776   & 0.9810   & 0.9758 \\ 
$\omega_b$                 & 0.02282  & 0.02296 & 0.02264 \\ 
$\omega_{cdm}$             & 0.1215   & 0.1229   & 0.1168 \\ 
$\tau_{\rm reio}$          & 0.06433  & 0.06619  & 0.06561 \\ 
$\ln(10^{10}A_s)$          & 3.0577    & 3.0639    & 3.0648 \\ 
$\Omega_m$                 & 0.295   & 0.297   & 0.297 \\ 
$S_8$                      & 0.8107   & 0.8095   & 0.8064 \\ 
$H_0$ [km/s/Mpc]           & 70.04    & 70.63    & 68.71 \\ 
\hline
$\chi^2_{\rm CMB}$         & 822.2    & 824.2    & 815.5 \\
$\chi^2_{\rm BAO}$         & 10.4    & 10.6   & 10.5 \\
$\chi^2_{\rm SN}$          & 1407.4   & 1406.7  & 1407.3 \\
$\chi^2_{\rm Mb}$          & 15.6    & 10.8    & 29.7 \\
\hline
$\chi^2_{\rm tot}$         & 2221.5  & 2216.1   & 2226.1 \\ 
\hline
\end{tabular}
\caption{Best-fit cosmological parameters for WZDR, nuADaM, and $\Lambda$CDM models with SH0ES prior. $\chi^2_\text{CMB}$ includes the contributions from {\bf P-ACT} and CMB lensing.}
\label{tab:best-fit}
\end{table}

An important consideration when interpreting constraints from high-$\ell$ CMB data concerns the contribution of secondary anisotropies, primarily gravitational lensing, which become large at small scales~\cite{McCarthy:2021lfp,Trendafilova:2025dce}. This introduces sensitivity to the modeling of non-linear physics, which can be affected by the dynamics of the models under consideration. While the ACT DR6 analysis extends to $\ell \sim 6500$ with foreground modeling, these secondary sources become dominant around $\ell \gtrsim 3000$. To assess the multipole range that is providing the dominant constraining power on dark radiation models and consequently verify the robustness of the results against potential sensitivity to modeling of non-linear scales, we conducted an additional MCMC analysis restricting ACT data to $\ell_{\rm max} = 3000$ while maintaining the standard Planck coverage at lower multipoles.

\begin{figure}[h]
\centering
    \includegraphics[width=0.49\textwidth]{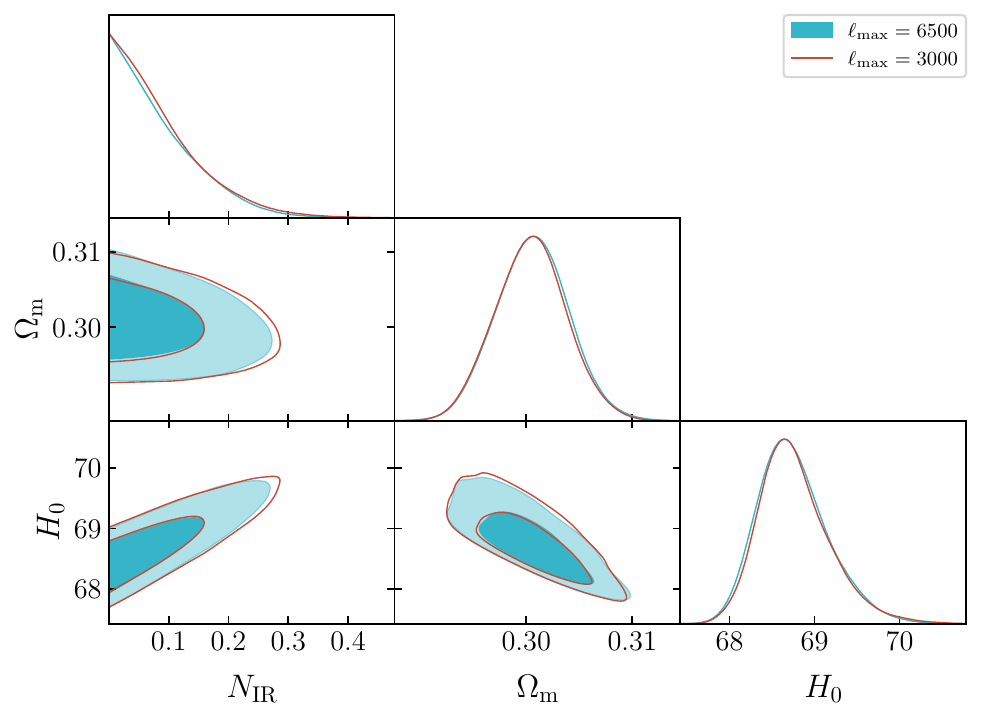}
    \includegraphics[width=0.49\textwidth]{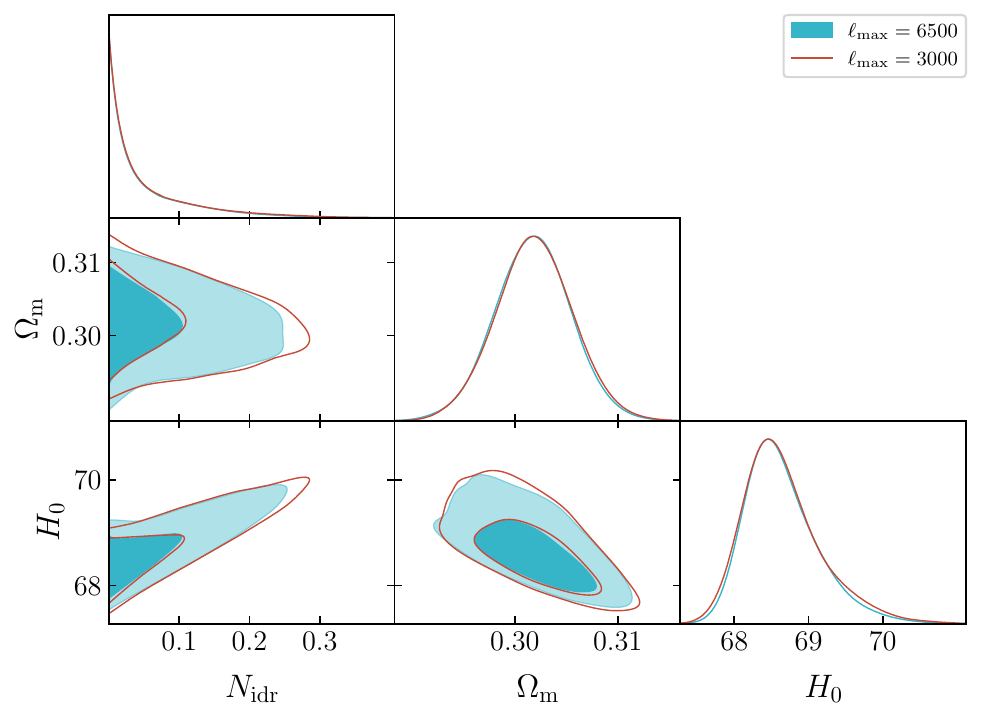}
\caption{WZDR~(left) and nuADaM~(right) fit to the full $\ell_\mathrm{max}=6500$ range of ACT DR6 in P-ACT-LBS and the truncated P-ACT-LBS$|_{\ell<3000}$.}
\label{fig:lmax3000}
\end{figure}

We find that the WZDR and nuADaM constraints remain essentially unchanged when limiting ACT to $\ell < 3000$, as can be seen in Fig.~\ref{fig:lmax3000}. For WZDR with P-ACT-LBS$|_{\ell < 3000}$, we obtain $N_{\rm IR} < 0.266$ (95\% CL), compared to $N_{\rm IR} < 0.246$ (95\% CL) with the full $\ell_\mathrm{max} = 6500$ range, representing only a small degradation in constraining power despite excluding half the multipole range. Similarly, nuADaM yields $N_{\rm idr} < 0.256$ (95\% CL) with $\ell_{\rm max} = 3000$ versus $N_{\rm idr} < 0.237$ (95\% CL) with full coverage. This robustness demonstrates that ACT's constraints on dark radiation arise primarily from the intermediate damping tail regime ($1000 < \ell < 3000$) where Silk damping effects are cleanly measurable and secondary anisotropies remain subdominant, rather than from the very highest multipoles where secondaries and their modeling framework, calibrated for $\Lambda$CDM, could introduce additional systematic biases when applied to models with non-$\Lambda$CDM physics.

To further investigate the nature of potential extra radiation, we conducted an analysis allowing both free-streaming and self-interacting (fluid-like) dark radiation components to vary independently. This two-parameter extension, parameterized by $N_{\rm fs}$ (free-streaming radiation) and $N_{\rm idr}$ (interacting dark radiation forming a perfect relativistic fluid with $w = c_s^2 = 1/3$), tests whether ACT DR6's preference for total radiation below the Standard Model expectation depends on the perturbative behavior of the extra radiation. While free-streaming and self-interacting radiation experience Silk damping identically to standard neutrinos, perturbations in self-interacting dark radiation propagate as acoustic waves, leading to slight increase in the amplitudes of perturbations and produces smaller phase shifts in CMB power spectra, potentially accommodating larger radiation densities. For this fits we used DESI DR1~\cite{DESI:2024mwx} data instead of DR2, to be able to more directly compare with the results in Ref.~\cite{ACT:2025tim}.

\setlength{\tabcolsep}{8pt}
\renewcommand{\arraystretch}{1.3}
\begin{table}[ht]
\centering
\begin{tabular}{|l|c|c|c|}
\hline
WZDR params & p-ACT-LBS & p-ACT-LBS+SH0ES & p-ACT-LBS$|_{\ell<3000}$ \\
\hline

$N_{\mathrm{IR}}$
& $< 0.246$
& $0.3147^{+0.221}_{-0.219}$
& $< 0.266$ \\

$\log_{10}a_{\mathrm{t}}$
& $-4.484^{+0.464}_{-0.495}$
& $-4.506^{+0.484}_{-0.474}$
& $-4.496^{+0.476}_{-0.482}$ \\

$\omega_{\mathrm{cdm}}$
& $0.11879^{+0.00254}_{-0.00188}$
& $0.12152^{+0.00341}_{-0.00347}$
& $0.11880^{+0.00276}_{-0.00190}$ \\

$n_s$
& $0.9747^{+0.00593}_{-0.00564}$
& $0.9778^{+0.00610}_{-0.00618}$
& $0.9751^{+0.00573}_{-0.00598}$ \\

$100\,\theta_s$
& $1.04182^{+0.00047}_{-0.00045}$
& $1.04200^{+0.00049}_{-0.00048}$
& $1.04182^{+0.00049}_{-0.00048}$ \\

$\omega_b$
& $0.02262^{+0.00023}_{-0.00021}$
& $0.02283^{+0.00025}_{-0.00024}$
& $0.02262^{+0.00023}_{-0.00022}$ \\

$\tau_{\mathrm{reio}}$
& $0.0638^{+0.0118}_{-0.0112}$
& $0.0660^{+0.0134}_{-0.0115}$
& $0.0638^{+0.0127}_{-0.0113}$ \\

$\ln(10^{10}A_s)$
& $3.0602^{+0.0214}_{-0.0203}$
& $3.0605^{+0.0241}_{-0.0215}$
& $3.0594^{+0.0223}_{-0.0199}$ \\

$\Omega_m$
& $0.3006^{+0.00678}_{-0.00649}$
& $0.2949^{+0.00656}_{-0.00641}$
& $0.3004^{+0.00676}_{-0.00651}$ \\

$H_0\,[\mathrm{km\,s^{-1}\,Mpc^{-1}}]$
& $68.75^{+0.92}_{-0.71}$
& $70.13^{+1.10}_{-1.08}$
& $68.77^{+0.97}_{-0.72}$ \\

\hline
\end{tabular}

\caption{Means with two-sigma (2$\sigma$) uncertainties for WZDR fits to three datasets. The first two are described in Sec.~\ref{sec:data&m}, while the third is the same as the first but restricting the likelihood to $\ell \leq 3000$ as described in the text.}
\end{table}

\setlength{\tabcolsep}{8pt}
\renewcommand{\arraystretch}{1.3}

\begin{table}[ht]
\centering
\begin{tabular}{|l|c|c|c|}
\hline
nuADaM params & p-ACT-LBS & p-ACT-LBS+SH0ES & p-ACT-LBS$|_{\ell<3000}$ \\
\hline
$N_{\mathrm{idr}}$
& $< 0.237$
& $0.3322^{+0.203}_{-0.201}$
& $< 0.256 $ \\

$\log_{10}(m_e/m_p)_{\mathrm{dark}}$
& $-3.946^{+0.420}_{-0.509}$
& $-3.851^{+0.306}_{-0.558}$
& $-3.932^{+0.405}_{-0.522}$ \\

$r_{\mathrm{twin}}$
& $0.0187^{+0.0692}_{-0.0183}$
& $0.0116^{+0.0187}_{-0.0108}$
& $0.0208^{+0.0718}_{-0.0204}$ \\

$\omega_{\mathrm{cdm}}$
& $0.11672^{+0.00480}_{-0.00929}$
& $0.12246^{+0.00398}_{-0.00380}$
& $0.11655^{+0.00537}_{-0.00960}$ \\

$n_s$
& $0.9759^{+0.00675}_{-0.00623}$
& $0.9807^{+0.00723}_{-0.00748}$
& $0.9759^{+0.00721}_{-0.00698}$ \\

$100\,\theta_s$
& $1.04191^{+0.00056}_{-0.00051}$
& $1.04235^{+0.00063}_{-0.00063}$
& $1.04192^{+0.00057}_{-0.00052}$ \\

$\omega_b$
& $0.02261^{+0.00025}_{-0.00022}$
& $0.02289^{+0.00027}_{-0.00028}$
& $0.02261^{+0.00026}_{-0.00023}$ \\

$\tau_{\mathrm{reio}}$
& $0.0650^{+0.0130}_{-0.0116}$
& $0.0667^{+0.0137}_{-0.0121}$
& $0.0643^{+0.0124}_{-0.0112}$ \\

$\ln(10^{10}A_s)$
& $3.0637^{+0.0231}_{-0.0221}$
& $3.0642^{+0.0249}_{-0.0226}$
& $3.0625^{+0.0230}_{-0.0200}$ \\

$\Omega_m$
& $0.3018^{+0.00682}_{-0.00689}$
& $0.2968^{+0.00715}_{-0.00660}$
& $0.3019^{+0.00711}_{-0.00690}$ \\

$H_0\,[\mathrm{km\,s^{-1}\,Mpc^{-1}}]$
& $68.64^{+1.15}_{-0.69}$
& $70.48^{+1.18}_{-1.22}$
& $68.65^{+1.21}_{-0.75}$ \\

\hline
\end{tabular}

\caption{Means with $2\sigma$ uncertainties for nuADaM fits to three datasets. The first two are described in Sec.~\ref{sec:data&m}, while the third is the same as the first but restricting the likelihood to $\ell \leq 3000$ as described in the text.}
\end{table}

\begin{figure}[h]
\centering
\includegraphics[width=0.5\textwidth]{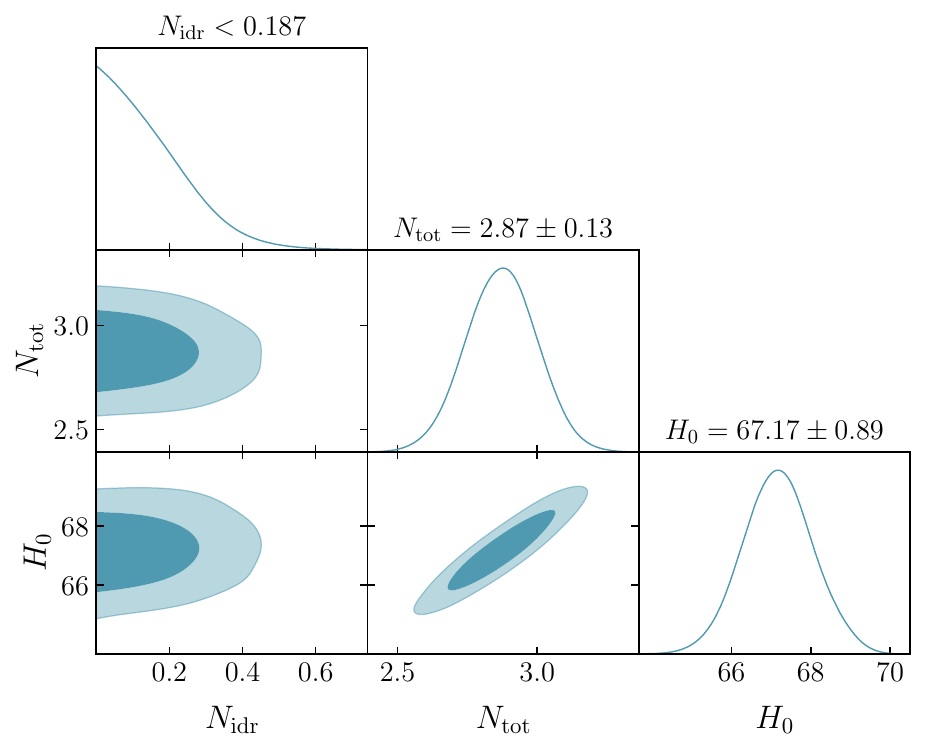}
\caption{Generalized dark radiation model containing a free-streaming component (which include the SM neutrinos) parametrized by \Neff and a self-interacting componenet parametrized by $N_{\rm idr}$. The total amount of energy density in radiation is parametrized by $N_{\rm tot} = N_{\rm eff} + N_{\rm idr}$.}
\label{fig:nfld_noSH0ES}
\end{figure}

We find that the total effective number of relativistic species remains consistent with (or slightly below) the Standard Model expectation regardless of the radiation's perturbative nature. With P-ACT-LBS, the total radiation density $N_{\rm tot} = N_{\rm fs} + N_{\rm idr}$ yields $N_{\rm tot} = 2.87 \pm 0.13$ (68\% CL), statistically indistinguishable from the baseline $\Lambda$CDM+$N_{\rm eff}$ result of $N_{\rm eff} = 2.86 \pm 0.13$ reported in the ACT DR6 baseline cosmology paper~\cite{ACT:2025fju}. The data show no preference for partitioning extra radiation between free-streaming and interacting components; the posterior for $N_{\rm idr}$ remains consistent with zero. This demonstrates that ACT DR6's slight preference for lower total radiation is robust to assumptions about dark radiation interactions, and that the high-$\ell$ temperature and polarization data provide decisive constraints on the energy density of extra radiation independent of its clustering properties. Consequently, the stringent bounds on  WZDR and nuADaM reflect more general restrictions on early-universe radiation density rather than sensitivity to the specific scale-dependent signatures of these models.

This result has implications for interpreting ACT DR6 constraints in the broader context of cosmological tensions. The constraining power of intermediate-scale damping tail measurements ($\ell \sim 1000-3000$) suggests that future improvements in small-scale CMB observations at $\ell > 3000$, while valuable for other science goals, will provide limited additional constraining power on early-universe radiation density beyond what ACT DR6 already achieves. Conversely, future ground-based experiments like the Simons Observatory, which will achieve substantial improvements in noise at $\ell \sim 1000 - 3000$, should yield correspondingly tighter sensitivity to extra radiation models, potentially reaching sensitivity levels where even subtle deviations from $\Lambda$CDM in the radiation sector become detectable or definitively excluded.

\section{Conclusion}
We have analyzed the WZDR and nuADaM models using ACT DR6 data combined with Planck, CMB lensing, BAO, and supernova measurements. These models introduce dynamical effects in the dark radiation sector through a mass threshold transition, generating scale-dependent modifications to CMB observables. Previous analyses with Planck data found that both models could accommodate larger values of $\Delta N_{\rm eff}$ compared to simpler dark radiation scenarios, suggesting a promising direction to resolve the Hubble tension.

Our results demonstrate that ACT DR6's improved high-$\ell$ measurements impose stringent constraints on both models. When fit to P-ACT-LBS data without the SH0ES prior, we find $N_{\rm IR} < 0.246$ (95\% CL) for WZDR and $N_{\rm idr} < 0.237$ (95\% CL) for nuADaM, with mean values of $H_0 = 68.75$ km/s/Mpc and $H_0 = 68.64$ km/s/Mpc respectively. Including the SH0ES prior raises the inferred Hubble constant to $H_0 \sim 70$ km/s/Mpc for both models, but this comes at the cost of increased tension with the CMB data. These constraints are only slightly weaker than those on $\Lambda$CDM+$\Delta N_{\rm idr}$ and SIDR models.

To verify that these constraints arise from the damping tail rather than smaller scales which are more sensitive to non-linear physics, we performed fits restricting ACT data to $\ell_{\rm max} = 3000$. The resulting bounds remain nearly unchanged, demonstrating that ACT's constraining power on extra radiation comes primarily from the intermediate damping regime ($1000 < \ell < 3000$) where secondary anisotropies are subdominant.

Our analysis establishes that the scale-dependent modifications introduced by WZDR and nuADaM transitions are insufficient to accommodate significant extra radiation when confronted with ACT DR6's precision measurements of the high-$\ell$ damping tail. The enhanced Silk damping from additional relativistic species represents a significant observational hurdle that radiation-based solutions to the Hubble tension must overcome. While the models analyzed here do not successfully navigate this constraint, our results do not preclude all radiation-based solutions. It remains possible that alternative dark sector models could identify dynamics that introduce compensating effects through degeneracies with other cosmological parameters, potentially offsetting the damping tail suppression while maintaining agreement with high-precision CMB data. Such models would need to incorporate mechanisms distinct from the transitions in WZDR and nuADaM to achieve the required balance between raising $H_0$ and maintaining a good fit to CMB data. 

\section{Acknowledgments}

The authors thank Collin Hill, Joel Meyers, Diego Redigolo and Alexander van Engelen for useful conversations. The research of MJ and GMT is supported in part by the National Science Foundation under Grant Number PHY-2412828. The support and resources from the Center for High Performance Computing at the University of Utah are gratefully acknowledged. 

\bibliographystyle{JHEP}
\bibliography{act.bib}

\appendix
\section{Triangle plots}\label{app:triangle}

\begin{figure*}[h!]
	\centering
    \includegraphics[width=0.98\textwidth]{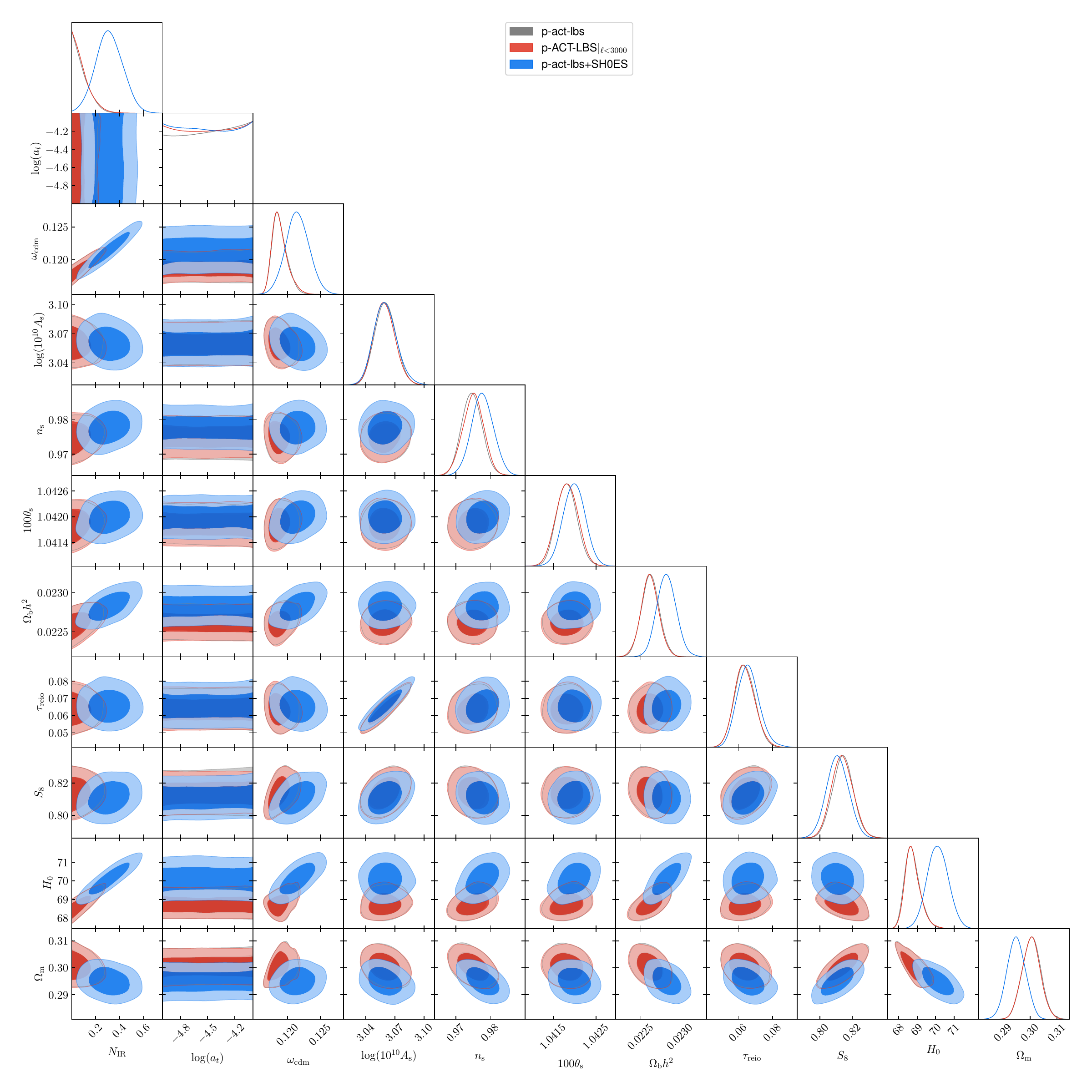}
    \caption{The WZDR fits to p-ACT-LBS, p-ACT-LBS$|_{\ell<3000}$, and p-ACT-LBS+SH0ES with all parameters.}
    \label{fig:wzdr-full-triangle}
\end{figure*}

\begin{figure*}[h!]
	\centering
    \includegraphics[width=0.98\textwidth]{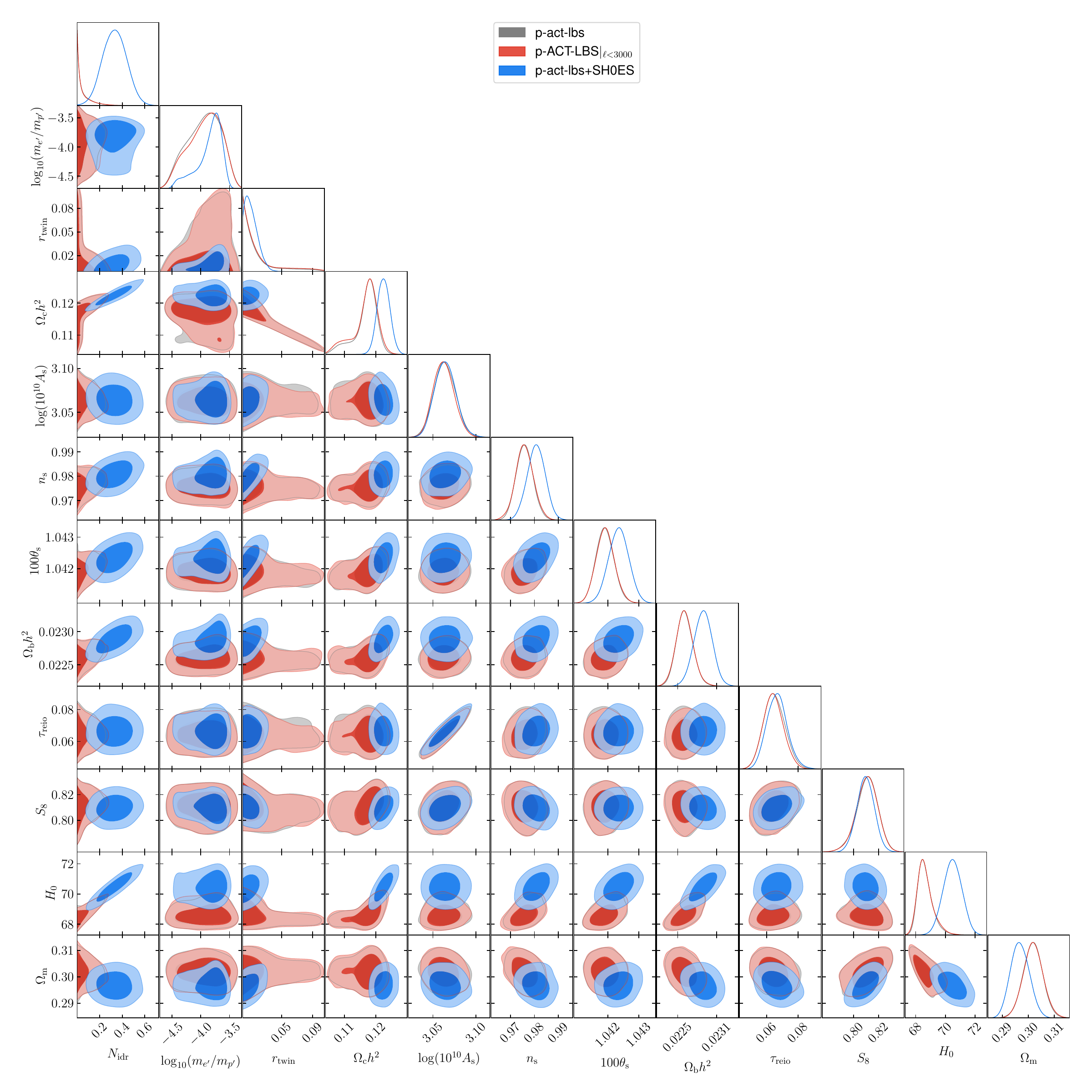}
    \caption{The nuADaM fits to p-ACT-LBS, p-ACT-LBS$|_{\ell<3000}$, and p-ACT-LBS+SH0ES }
    \label{fig:nuadam-full-triangle}
\end{figure*}

\end{document}